\documentclass[aps,twocolumn,superscriptaddress,showpacs,showkeys]{revtex4-2}
\usepackage{amsmath}
\usepackage{color,xcolor}
\usepackage{graphicx}
\newcommand{\bastar}{\begin{eqnarray*}}
\newcommand{\eastar}{\end{eqnarray*}}
\newskip\humongous \humongous=0pt plus 1000pt minus 1000pt

\newif\ifdtup

\relax
\newcommand{\W}{{\vec W}}
\newcommand{\n}{\hat n}

\newcommand{\hr}{\hat r}

\newcommand{\hD}{{\hat D}}
\newcommand{\bD}{{\bar D}}
\newcommand{\cD}{{\cal D}}
\newcommand{\pd}{\partial}

\newcommand{\hB}{{\hat B}}

\newcommand{\bA}{{\bar A}}

\newcommand{\tB}{{\tilde B}}
\newcommand{\tC}{{\tilde C}}

\newcommand{\bF}{{\bar F}}
\newcommand{\G}{{\vec G}}
\newcommand{\B}{{\vec B}}

\newcommand{\vPhi}{{\vec \Phi}}

\newcommand{\hG}{{\hat G}}

\newcommand{\mn}{{\mu\nu}}
\newcommand{\om}{\omega}

\newcommand{\lam}{\lambda}

\newcommand{\al}{\alpha}

\newcommand{\vap}{\varphi}

\newcommand{\be}{{\bar e}}

\newcommand{\vsig}{{\vec \sigma}}

\newcommand{\nn}{\nonumber}

\newcommand{\cL}{{\cal L}}

\begin{document}
\title{Non-Abelian Gauge Theory of Spin Triplet Superconductivity and Spin Triplet Magnon Spintronics}
\author{Franklin H. Cho}
\email{Contact author: cho.franklin@qns.science}
\affiliation{Center for Quantum Nano Science,
Ewha Woman's University, Seoul 03766, Korea}
\author{Y. M. Cho}
\email{Contact author: ymcho0416@gmail.com}
\affiliation{School of Physics and Astronomy,
Seoul National University, Seoul 08826, Korea}
\affiliation{Center for Quantum Spacetime, 
Sogang University, Seoul 04107, Korea}  
\author{Pengming Zhang}
\email{Contact author: zhangpm5@mail.sysu.edu.cn}
\affiliation{School of Physics and Astronomy,
Sun Yat-Sen University, Zhuhai 519082, China}
\author{Li-Ping Zou}
\email{Contact author: zoulp5@mail.sysu.edu.cn}
\affiliation{Sino-French Institute of Nuclear Engineering and Technology, Sun Yat-Sen University, Zhuhai 519082, China}

\begin{abstract}
We present an SU(2)xU(1) generalization of the Ginzburg-Landau theory of the spin triplet ferromagnetic superconductivity which could also describe the physics of the spin triplet magnon spintronics, where the SU(2) gauge interaction of the magnon plays an important role. The theory is made of the massive photon, massless neutral magnon, massive non-Abelian magnon, and the Higgs scalar field which represents the density of the Cooper pair. It has the following characteristic features, the long range magnetic interaction mediated by the massless magnon, two types of conserved supercurrents (the ordinary charge current and the spin current of the magnons), and the non-Abelian Meissner effect generated by the spin current. It has two types of vortices, the quantized magnetic vortex and spin vortex. Moreover, it has two types of monopoles, the monopole which has the ordinary magnetic charge and the one which has the spin magnetic charge. The theory is characterized by three scales. In addition to the correlation length fixed by the mass of the Higgs field it has two different mass scales, the one fixed by the mass of the photon and the other fixed by the mass of the off-diagonal magnon. We compare the theory with the non-Abelian gauge theory of the spin doublet ferromagnetic superconductivity which could also be interpreted as an effective theory of the electron spintronics. We discuss 
the physical implications of the non-Abelian 
gauge theories in condensed matter physics.
\end{abstract}


\maketitle

\section{Introduction}

The Ginzburg-Landau theory, the effective theory of the Abelian superconductivity, has helped us to understand the physics of 
the superconductivity very much \cite{lg}. 
It has allowed us to understand the nature of the supercurrent in terms of the Cooper pair and explained the Meissner effect with 
the supercurrent with two scales, 
the correlation length of the electron pairs fixed by the Higgs mass and the penetration length of the magnetic field fixed by the mass of the photon. Moreover, it has successfully explained the existence of the quantized magnetic vortex in terms of the Abrikosov vortex \cite{abri,no}.
 
But the advent of the non-Abelian superconductors made of the spin doublet and/or triplet Cooper pair motivates us to think of the possibility of a non-Abelian Ginzburg-Landau theory of superconductivity \cite{exp1,exp2}. This is because the spin doublet and/or triplet Cooper pair could naturally be identified as non-Abelian spin multiplet. For example the spin triplet Cooper pair made of two electron spins could be viewed as an SU(2) spin triplet. If so, one might ask if we could construct an effective non-Abelian gauge theory of superconductivity which could replace the Abelian Ginzburg-Landau theory. 

There have been efforts to generalize the Ginzburg-Landau theory to non-Abelian Ginzburg-Landau theory of superconductivity \cite{prb05,prb06,epjb08}. But in these earlier theories the genuine non-Abelian gauge interaction was not introduced, because such non-Abelian gauge interaction was not accepted in the condensed matter physics. This situation has changed, however, and the non-Abelian gauge interaction is being introduced in condensed matter physics these days. 

This allowed us to introduce the non-Abelian gauge interaction in superconductivity.
Indeed recently a non-Abelian gauge theory 
of two-gap ferromagnetic superconductivity, 
an SU(2)xU(1) extension of the Abelian Ginzburg-Landau theory, which could be viewed as an effective theory of two-gap ferromagnetic superconductors made of spin-up and spin-down Cooper pair doublet has been proposed \cite{pla23,ap24,arx250}. Just like 
the Abelian Ginzburg-Landau theory it has 
the U(1) electromagnetic interaction, but 
the new ingredient is the SU(2) magnon gauge interaction between the spin-up and spin-down Cooper pairs. A remarkable feature of the theory is the photon-magnon mixing between 
the U(1) potential and the diagonal part of the magnon potential. After the mixing it has massless neutral magnon and massive doubly charged magnon, in addition to the massive photon.  

So, just as in the ordinary Ginzburg-Landau theory, the massive photon generates the Meissner effect which screens the magnetic field. But unlike the ordinary Ginzburg-Landau theory, it has a long range magnetic interaction mediated by the massless magnon. Moreover, it has the doubly charged (the electric charge and the spin charge) massive magnon which is absent in ordinary superconductors, which generates 
a non-Abelian Meissner effect which screens 
the spin flux. The theory is characterized by three scales. In addition to the correlation length fixed by the mass of the Higgs field and the penetration lengths of the magnetic field fixed by the photon mass, it has a third scale fixed by the mass of the doubly charged magnon which is responsible for the non-Abelian Meissner effect. Moreover, recently this theory of two-gap ferromagnetic superconductivity has been generalized to 
the non-Abelian gauge theory of spin triplet superconductivity \cite{arx250}. 

The purpose of this paper is two-fold. 
The first purpose is to discuss the non-Abelian Ginzburg-Landau theory of spin triplet superconductivity made of the spin triplet Cooper pair in more detail. Just like the theory of spin doublet Cooper pair the new theory has the SU(2) magnon gauge interaction among spin triplet Cooper pairs, in addition to the U(1) electromagnetic interaction. So the theory becomes very much like the SU(2)xU(1) Ginzburg-Landau theory of spin doublet Cooper pair \cite{pla23,ap24,arx250}. The main difference is that here the Cooper pair becomes a spin triplet, and that there is no photon-magnon mixing. 

One might ask why do we need the SU(2) gauge interaction in the non-Abelian superconductors, when we could still obtain 
an interesting theory of the non-Abelian superconductivity treating the SU(2) symmetry a global symmetry \cite{prb05,prb06,epjb08}. Indeed, a genuine non-Abelian gauge interaction has rarely been discussed in condensed matter physics till recently, in spite of the fact that there have been huge amount of discussions on non-Abelian condensed matters in the literature \cite{lu}. 

There are two motivations for the non-Abelian gauge interaction in spin triplet superconductors. First, we could always switch off the gauge interaction if we like, and make the SU(2) a global symmetry \cite{prb05,
prb06,epjb08}. But the global symmetry is likely to become a local (i.e., gauge) symmetry under small perturbations of the symmetry. So it is natural to introduce the gauge interaction to the spin triplet Cooper pair. The second motivation is that recently there have been increasing evidence for the non-Abelian gauge interaction in condensed matter physics. We have two examples. First, a non-Abelian gauge interaction has been proposed to explain two-gap ferromagnetic superconductors \cite{pla23,ap24,arx250}. Second, similar non-Abelian gauge interactions have been proposed to explain the magnetism in frustrated magnetic materials recently \cite{arx250,zarkim}. This justifies us to introduce the non-Abelian gauge interaction 
in spin triplet superconductors. 

Of course, ultimately only the experiment could tell if there exists a gauge interaction among the spin triplet Cooper pairs or not. But theoretically the local gauge interaction seems a natural way to describe the spin triplet superconductors. 

From the physical point of view the generic features of the spin triplet Ginzburg-Landau theory become quite similar to the spin doublet Ginzburg-Landau theory \cite{pla23,
ap24,arx250}. For example, in both cases we have the massive photon which generates the ordinary Meissner effect and a long range interaction mediated by the massless magnon. In addition, we have the massive magnon which carries the spin charge $g'$. As importantly, we have two conserved supercurrents, the electromagnetic current and the spin current made of the magnons. 

The main difference is the photon-magnon mixing which exists in the spin doublet Ginzburg-Landau theory but is absent in the spin triplet superconductors. Because of this we have no mixing between the two conserved currents, and thus no conversion between the two currents, in the spin triplet superconductors. This makes the spin triplet Ginzburg-Landau theory simpler than the two-gap ferromagnetic Ginzburg-Landau theory.   

The other purpose of this paper is to show that exactly the same theory could also be interpreted as an effective theory of the spin triplet magnon spintronics. It has been well known that the magnon spintronics is closely related to the non-Abelian superconductivity \cite{chumak,jmmm,hb}, which has made the superconducting spintronics an important part of spintronics \cite{lind,esch}. In both cases the non-Abelian spin structure and the magnon spin-spin interaction play the central role, and the long range magnetic order becomes an essential feature of the theories. But these 
results were mostly the experimental observations, and there have been no theoretical explanations why this is so.

Recently, however, a theoretical explanation  was provided on why the theory of two-gap ferromagnetic superconductivity is closely connected to the theory of the electron spintronics \cite{arx250,arx251}. In this paper we show that this connection between 
the magnon spintronics and non-Abelian superconductivity is not just closely related, but in fact almost one to one for the spin triplet superconductors. This is because 
the spin triplet Cooper pair could also be interpreted as the spin triplet spintronic matter. With this interpretation the theory of the spin triplet superconductivity could describe 
the spin triplet spintronic materials, 
with the SU(2) gauge potential as 
the non-Abelian magnon and the U(1) gauge field as the electromagnetic potential. 
And just like the spin triplet superconductivity, the theory has 
the massless magnon which assures the long range magnetic order and the doubly charged massive magnon. Moreover, the theory has conserved charge and spin currents.
This is remarkable.

The paper is organized as follows. In Section II we discuss the non-Abelian Ginzburg-Landau theory of the spin-triplet superconductivity and discuss the physical properties of the theory. In Section III 
we discuss the topological objects of 
the theory, in particular the non-Abelian quantized magnonic vortex which carries the quantized spin flux $2\pi n/g'$ and magnonic monopole which has the spin magnetic charge $4\pi/g'$. In Section IV we review 
the non-Abelian Ginzburg-Landau theory 
of the two-gap ferromagnetic superconductivity to compare this with the theory of the spin triplet non-Abelian superconductivity. In Section V we discuss the topological objects in two-gap ferromagnetic superconductors. In Section VI we discuss the deep connection which could exist between this theory and the theory of the spin-triplet spintronics. Finally, in Section VII we discuss the implications of the non-Abelian gauge theories in condensed matter physics. 

\section{Non-Abelian gauge theory of spin triplet superconductivity}

In non-Abelian ferromagnetic superconductors 
we can have a spin triplet Cooper pair made of an
electron pair which forms a spin triplet. In this case we need a non-Abelian gauge theory 
of the spin triplet ferromagnetic superconductivity. To construct such a theory we let $\vPhi=(\phi_1,\phi_2,\phi_3)$ the complex spin triplet Cooper pair which forms an SU(2) triplet, and consider the following SU(2)xU(1) gauge theory described by 
the Lagrangian \cite{arx250}
\begin{gather}
\cL =-\frac12 |\cD_\mu \vPhi|^2 -V(\vPhi)
-\frac14 F_\mn^2-\frac14 \G_\mn^2, \nn \\
\cD_\mu \vPhi =(D_\mu -i g A_\mu) \vPhi,
~~~D_\mu \vPhi=(\pd_\mu + g' \B_\mu \times \big) \vPhi,   \nn\\
V(\Phi) =\frac{\lambda}{2}\big(|\vPhi|^2
-\frac{\mu^2}{\lambda}\big)^2,
\label{3lag0}
\end{gather}
where $A_\mu$ and $\B_\mu$ are the ordinary electromagnetic U(1) and the SU(2) gauge potentials which describe the photon and  magnons, $F_\mn$ and $\G_\mn$ are the corresponding field strengths, $g$ and $g'$ 
are the coupling constants, and $V(\vPhi)$ is the self-interaction potential of the Cooper pair which we assume to have the above quartic form in this paper for simplicity. For the Cooper pair we have $g=2e$, but for the moment we leave $g$ and $g'$ arbitrary. 

Notice that this Lagrangian reduces to the ordinary Ginzburg-Landau Lagrangian if we neglect the spin structure of the Cooper pair and remove the magnon interaction.
This tells that the Lagrangian (\ref{3lag0}) is a natural non-Abelian extension of the Abelian Ginzburg-Landau theory.

To understand the physical meaning of 
the non-Abelian magnon interaction, we need 
the Abelian decomposition of the non-Abelian 
magnon potential $\B_\mu$. Let $(\n_1,\n_2,\n_3)$ be an arbitrary SU(2) orthonormal frame, and choose an arbitrary direction $\n$ as the Abelian direction and let $\n=\n_3$. With this we can decompose $\B_\mu$ to the restricted potential $\hB_\mu$ and the valence potential $\W_\mu$ gauge independently \cite{ap24,prd80,prl81},
\begin{gather}
\B_\mu= \hB_\mu +\W_\mu,  \nn\\
\W_\mu =W_\mu^1 \n_1 +W_\mu ^2 \n_2,~~~\n \cdot \W_\mu=0,  \nn\\
\hB_\mu= B_\mu \n -\frac1{g'} \n \times \pd_\mu \n=\tB_\mu +\tC_\mu,  \nn\\
\tB_\mu= B_\mu \n~~~~(B_\mu=\n \cdot \B_\mu), \nn\\
\tC_\mu=-\frac{1}{g'} \n\times \pd_\mu \n,
\nn\\
\hG_\mn= \pd_\mu \hB_\nu-\pd_\nu \hB_\mu
+ g' \hB_\mu \times \hB_\nu =G_\mn' \n, \nn \\
G'_\mn=G_\mn + H_\mn
= \pd_\mu B_\nu'-\pd_\nu B_\mu', 
\label{cdec}
\end{gather}
where $\hB_\mu$ is the Abelian projection of 
$\B_\mu$ fixed by the isometry condition
\begin{gather}
D_\mu \n =0.
\label{ap}
\end{gather}
Notice that $\hB_\mu$ is the potential which makes the Abelian direction a covariant constant, the potential which parallelizes $\n$.

With this Abelian decomposition the Lagrangian (\ref{3lag0}) can be expressed by
\begin{gather}
\cL =-\frac12 |\cD_\mu \vPhi|^2 -V(\vPhi)
-\frac14 F_\mn^2-\frac14 \hG_\mn^2 \nn \\
-\frac14 (\hD_\mu\W_\nu-\hD_\nu\W_\mu)^2
-\frac{g'}{2} \hG_\mn \cdot (\W_\mu \times \W_\nu) \nn \\
-\frac{g'^2}{4} (\W_\mu \times \W_\nu)^2,  \nn\\  
D_\mu \vPhi=(\hD_\mu + g' \W_\mu \times \big) \vPhi, \nn\\
\hD_\mu = \pd_\mu + g' \hB_\mu \times.  
\label{3lag1}
\end{gather}
To proceed, let us express $\vPhi$ by the scalar field $\rho$ which represents the density of 
the electron Cooper pair and the unit SU(2) triplet $\n$ which represents the spin degrees of the Cooper pair by
\begin{gather}
\vPhi= \rho~\exp (-i\theta)~\n,
~~~\rho=|\vPhi|,~~~\n^2 = 1.
\label{Phi}
\end{gather}
With this we have 
\begin{gather}
\cD_\mu \vPhi =\Big[\big(\pd_\mu \rho 
-ig \rho \bA_\mu \big)~\n  
+g' \rho~\W_\mu \times \n \Big] \exp(-i\theta),   \nn\\ 
\bA_\mu =A_\mu +\frac1g \pd_\mu \theta,
\end{gather}
so that the Lagrangian can be expressed by
the physical fields
\begin{gather}
\cL =- \frac12 (\pd_\mu \rho)^2 
-\frac{\lambda}{2}\big(\rho^2 -\rho_0^2 \big)^2
-\frac14 \bF_\mn^2 
-\frac{g^2}{2} \rho^2 \bA_\mu^2   \nn\\
-\frac14 {G'}_\mn^2    
-\frac14 (\hD_\mu\W_\nu-\hD_\nu\W_\mu)^2
-\frac{g'^2}{2} \rho^2 \W_\mu^2  \nn\\
-\frac{g'}{2} G'_\mn \n \cdot (\W_\mu \times \W_\nu)
-\frac{g'^2}{4} (\W_\mu \times \W_\nu)^2.
\label{3lag2}
\end{gather}
This can also be expressed in an Abelianized form in the complex notation by
\begin{gather}
\cL=- \frac12 (\pd_\mu \rho)^2 
-\frac{\lambda}{2}\big(\rho^2 
-\rho_0^2 \big)^2  
-\frac14 \bF_\mn^2 
-\frac{g^2}{2} \rho^2 \bA_\mu^2  \nn\\
-\frac14 {G'}_\mn^2   
-\frac12 |D'_\mu W_\nu -D'_\nu W_\mu|^2
- g'^2 \rho^2 W_\mu^* W_\mu  \nn\\
+ig' G'_\mn W_\mu^* W_\nu
+\frac{g'^2}{4} (W_\mu^*W_\nu
-W_\nu^*W_\mu)^2, \nn \\  
D_\mu'=\pd_\mu +ig' B_\mu',  
~~~W_\mu =\frac{1}{\sqrt 2} (W^1_\mu +i W^2_\mu).
\label{3lag3}
\end{gather}
Notice that the Lagrangians (\ref{3lag2}) and (\ref{3lag3}) are simple re-parametrizations of the Lagrangian (\ref{3lag0}), so that they are mathematically identical to the original one. So, they inherit the original SU(2)xU(1) gauge symmetry of (\ref{3lag0}). Moreover, here the spin triplet Cooper pair $\vPhi$ has completely disappeared. The only remaining part is 
the density of the Cooper pair $\rho$, and 
the other three degrees of $\vPhi$ have been absorbed to the photon $A_\mu$ and magnon $\W_\mu$ making them massive. This of course is the mass generation by the Higgs mechanism. 

However, here this mass generation is coming from the non-vanishing vacuum density of 
the electron pair $\rho_0$, not by any spontaneous symmetry breaking of the spin triplet Cooper pair. This is because we have $\langle \vPhi \rangle= \rho_0 ~\exp{(i \langle \theta \rangle)}~\langle \n \rangle$, so that the orientation of the spin direction of the Cooper pair $\langle \n \rangle$ and the Abelian electromagnetic phase $\langle \theta \rangle$ can still fluctuate with the Higgs mechanism, even at the vacuum. This tells that the popular Higgs mechanism by spontaneous symmetry breaking is a misleading explanation which reflects only half of the full story \cite{pla23,ap24,arx250}.

The theory has the following important features. First of all, it has a long range magnetic order. This is because the massless magnon $B_\mu'$ generates a long range magnetic interaction. This type of the long range magnetic order is a characteristic feature common in the ferromagnetic materials, frustrated magnetic materials, and the spintronic materials \cite{pla23,ap24,arx250}.

Second, it has two conserved supercurrents, the electromagnetic current $J_\mu^{(e)}$ and the spin current $J_\mu^{(s)}$. This is because it has two U(1) gauge symmetries, the electromagnetic U(1) and the U(1) subgroup of SU(2) which leaves $\n$ invariant. To see this, notice that from (\ref{3lag3}) we have the following equations of motion
\begin{gather}
\pd_\mu^2\rho-g^2\rho \bar A_\mu^2 
-2 g'^2 \rho W_\mu^* W_\mu
=2\lambda (\rho^2-\rho_0^2) \rho,   \nn\\
D_\mu' (D_\mu' W_\nu -D_\nu' W_\mu) 
=g'^2 \rho^2 W_\nu +ig'G'_\mn W_\mu \nn\\
-g'^2 \Big((W_\mu^* W_\nu
-W_\nu^* W_\mu) W_\mu \Big),  \nn\\
\pd_\mu \bF_\mn =J_\nu^{(e)},  \nn\\
\pd_\mu G'_\mn =J_\nu^{(s)},  
\label{3eom}
\end{gather}
where
\begin{gather}
J_\nu^{(e)} =g^2\rho^2\bar{A}_\nu,\nn\\  
J_\nu^{(s)} =ig'\Big(W_\mu^*(D_\mu' W_\nu
-D_\nu' W_\mu) \nn\\
-W_\mu(D_\mu' W_\nu-D_\nu' W_\mu)^* \Big).
\label{3cJ}
\end{gather}
Obviously the last two equations tell that 
the theory has two conserved currents $J_\mu^{(e)}$ and $J_\mu^{(s)}$. 

Third, it has the non-Abelian Meissner effect
which screens the spin flux of the magnons, 
as well as the ordinary Meissner effect which screens the magnetic flux. This is because the spin current of the massive magnon acts against the spin flux to screen it, just as the electric supercurrent of the massive photon screens the magnetic flux in 
the ordinary superconductors. This must be clear from (\ref{3cJ}). This point will become evident when we discuss the quantized non-Abelian spin vortex in the following. 

There are three points to be mentioned before we leave this section. First, this is a theory in which the spin-spin interaction is described by the exchange of the messenger particles, the magnons. Traditionally the spin-spin interaction in physics has always been treated as an instantaneous action at a distance. Of course the magnon has been thought to be responsible for the spin-spin interaction, but till recently there have been few self-consistent field theories which describe the spin-spin interaction by the exchange of the magnons in terms of the Feynman diagrams \cite{pla23,ap24,arx250}. This has been strange, because in modern physics all fundamental interactions in nature are described by the exchange of the messenger particles. The above theory does exactly that. 

Second, it should be pointed out that 
the above theory should be regarded as a natural extension of the earlier works
mentioned before \cite{prb05,prb06,epjb08}.
In these earlier works a genuine non-Abelian gauge interaction was not considered because such interaction was considered as heretic 
in condensed matter physics. Now, the time has changed, and we can generalize these works to include the non-Abelian gauge interaction.   

Finally, this theory is deeply related to 
the theory originally proposed by Georgi and Glashow in an attempt to unify the weak and electromagnetic interactions in high energy physics \cite{gg}. In fact, when we remove 
the complex phase in the Higgs triplet $\vPhi$ and switch off the electromagnetic U(1) gauge interaction in the Lagrangian (\ref{3lag0}), it becomes exactly the Georgi-Glashow Lagrangian. This shows 
that our theory of the spin triplet superconductors is a straightforward generalization of the Georgi-Glashow model.

Moreover, this Georgi-Glashow Lagrangian 
is precisely the type of Lagrangian 
proposed recently to describe the magnetic interaction in frustrated magnetic 
materials in condensed matter physics \cite{arx250,zarkim}. This tells that our theory of the spin triplet superconductivity is also closely related to the theory of magnetism in frustrated magnetic materials. It is really remarkable that an almost forgotten unrealistic theory in high energy physics has come back to describe real non-Abelian physics in condensed matter physics. This point has an interesting implication in the following as we will 
see soon.  

\section{Topological Objects in Spin Triplet Superconductors}

The spin triplet superconductors have interesting topological objects. This is because the theory has two $\pi_1(S^1)$ string topologies coming from the electromagnetic U(1) and the Abelian U(1) subgroup of the SU(2), as well as the $\pi_2(S^2)$ monopole topology coming from the SU(2) gauge symmetry. 

\begin{figure}
\includegraphics[height=4.5cm,width=7.5cm]
{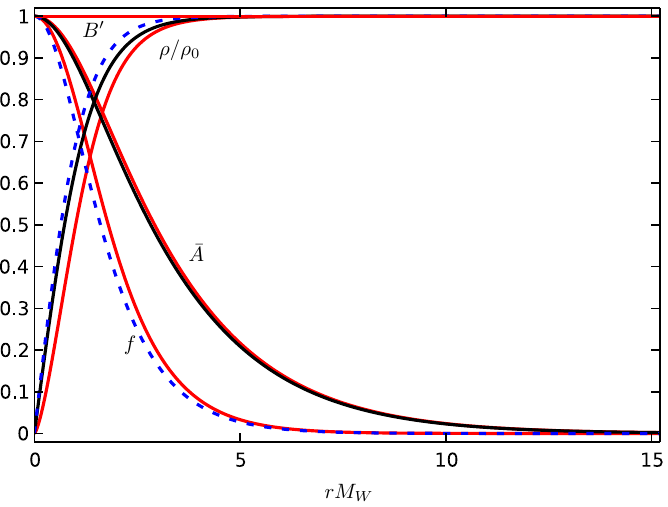}
\caption{\label{3scabv} The Abrikosov vortex in the spin triplet superconductors for $m=1$ is shown in black curves, and the Abrikosov vortex which has the massive magnon dressing (the $f$ profile) is shown in red curves. These solutions exist with or without the singular spin vortex which carries the spin flux $2\pi n/g'$, which is shown in red curve by $B'=1$ where we have put $g'=2g$. Notice that the spin vortex made of the Higgs scalar and massive magnon  shown in Fig. \ref{3scnav} is plotted in dotted blue curves for comparison.}
\end{figure}

To show this, we consider the quantized magnetic vortex of the Lagrangian (\ref{3lag0}) first, and choose the following vortex ansatz in the cylindrical coordinates $(r,\vap,z)$, 
\begin{gather}
\vPhi= \rho(r)~\exp (-im\varphi)~\n,   
~~\n= \left(\begin{array}{ccc}
\sin \al(r) \cos (n \vap) \\ 
\sin \al(r) \sin (n \vap) \\
\cos \al(r) \end{array} \right), \nn\\
A_\mu =\frac{m}{g} A(r)~\pd_\mu \vap, \nn\\ 
\hB_\mu=\frac{n}{g'} B(r) \pd_\mu \vap~\n -\frac{1}{g'}~\n \times \pd_\mu \n,  \nn\\
\W_\mu =\frac{1}{g'} f(r)
~\n \times \pd_\mu \n,
\label{3vans0}
\end{gather} 
where $m$ and $n$ are integers which represent the winding numbers of the $\pi_1(S^1)$ topology of U(1) and U(1) subgroup of SU(2). In terms of the physical field the ansatz becomes
\begin{gather}
\rho=\rho(r), 
~~~~\bA_\mu =\frac{m}{g} 
\big(A +1 \big)~\pd_\mu \vap
=\frac{m}{g} \bA~\pd_\mu \vap,  \nn\\
B_\mu' =-\frac{n}{g'} \big(1-\cos \al -B \big) ~\pd_\mu \vap 
=-\frac{n}{g'} B'~\pd_\mu \vap,   \nn\\
\W_\mu= \frac{f}{g'}~\n \times \pd_\mu \n.
\label{3vans1}
\end{gather}
This suggests that the ansatz could describe two types of vortices, the electromagnetic vortex described by $\bA_\mu$ and the spin vortex described by $B_\mu'$. 

\begin{figure}
\includegraphics[height=4.5cm,width=7.5cm]
{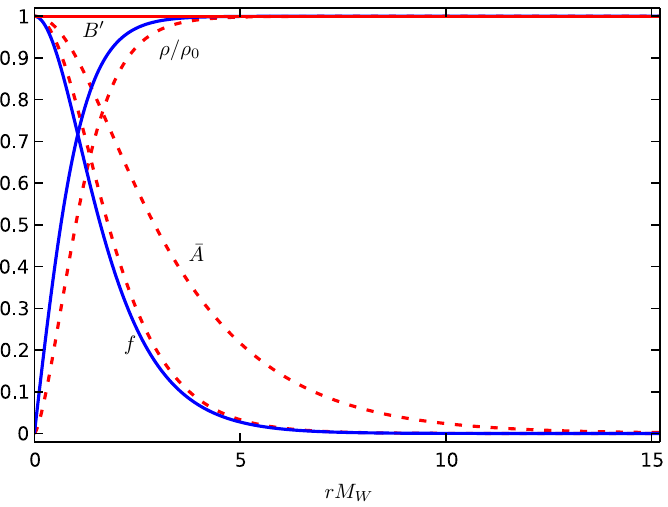}
\caption{\label{3scnav} The singular spin vortex solution with $B'=1$ in red line, which has the spin flux $2\pi n/g'$ which has the massive magnon and Higgs scalar dressing (for $n=1$) in blue curves, in spin triplet ferromagnetic superconductors. Notice that the Abrikosov vortex which has the massive magnon dressing shown in Fig. \ref{3scabv} is plotted in dotted red curves for comparison.}
\end{figure}

Assuming that $\al$ is a constant, we have the following equations of motion from the ansatz, 
\begin{gather}
\ddot \rho +\frac{\dot \rho}{r}
-\frac{m^2\bA^2+n^2 f^2 \sin^2 \al}{r^2} \rho
= 2 \lambda (\rho^2-\rho_0^2) \rho,  \nn\\
\ddot \bA-\frac{\dot \bA}{r}
=g^2 \rho^2 \bA,   \nn\\
\ddot B'-\frac{\dot B'}{r} =0,   \nn\\
\sin \al \Big(\ddot f +\frac{\dot f}{r} 
-g'^2 \rho^2 f \Big) =0,  \nn\\
n^2 \sin \al \Big(\dot{f}(B'-1) 
-f \dot B' \Big)=0.
\label{3veom}
\end{gather}
This has interesting solutions. 
First, when $\al=0$, the equation reduces to
\begin{gather}
\ddot \rho +\frac{\dot \rho}{r}
-\frac{m^2 \bA^2}{r^2}\rho
=2\lambda (\rho^2-\rho_0^2)~\rho,  \nn\\
\ddot \bA -\frac{\dot \bA}{r}
-g^2 \rho^2 \bA =0,  \nn\\
\ddot B'-\frac{\dot B'}{r} =0.
\label{3abe}
\end{gather}
So, when $B'=0$, this becomes the equation for the well known Abrikosov vortex. Solving this with the boundary condition
\begin{gather}
\rho(0)=0,~~~\rho(\infty)=\rho_0,
~~~\bA(0)=1,~~~\bA(\infty)=0,
\end{gather}
we obtain the Abrikosov vortex made of 
the Higgs scalar and massive photon
carrying quantized magnetic flux $2\pi m/g$,
\begin{gather}
\Phi=\oint_{r=0}^{r=\infty} \bA_\mu dx^\mu 
= -\frac{2\pi m}{g},~~~\bA_\mu =\frac{m}{g} \bA~\pd_\mu \vap.
\label{amf}
\end{gather}
This is shown in Fig. \ref{3scabv} in black curves. Moreover, we could also have  
the Abrikosov vortex with $B'=1$, 
so that the Abrikosov vortex can coexist with the singular spin vortex carrying 
the quantized spin flux $2\pi/g'$,
\begin{gather}	
\Phi=\oint B'_\mu dx^\mu 
=-\frac{2\pi n}{g'},~~~B_\mu' =-\frac{n}{g'} \pd_\mu \varphi.
\label{3ssv}	
\end{gather}   
This is shown in Fig. \ref{3scabv} in red line.

Second, with $\al=\pi/2$ and $B'=1$, (\ref{3veom}) reduces to
\begin{gather}
\ddot \rho+\frac{\dot \rho}{r}
-\frac{m^2 \bA^2+n^2f^2}{r^2}\rho
=2\lambda (\rho^2-\rho_0^2) \rho,\nn\\
\ddot \bA-\frac{\dot \bA}{r}
=g^2 \rho^2 \bA,   \nn\\
\ddot f-\frac{\dot f}{r}=g'^2 \rho^2 f.
\label{3ve2}
\end{gather}
We can solve this with $\bA=0$ and with 
the boundary condition
\begin{gather}
\rho(0)=0,~~~\rho(\infty)=\rho_0,
~~~f(0)=1,~~~f(\infty)=0,
\end{gather}
and obtain the quantized spin vortex solution (\ref{3ssv}) which has Higgs 
scalar and massive magnon dressing. 
This vortex solution is shown in Fig. \ref{3scnav} in blue curve.  

Moreover, we can solve (\ref{3ve2}) with 
a non-trivial $\bA$ imposing the boundary condition 
\begin{gather}
\rho(0)=0,~~~~\bA(0)=1,~~~~f(0)=1,  \nn\\
\rho(\infty)=\rho_0,~~~~\bA(\infty)=0,
~~~~f(\infty)=0,
\label{3vbc}
\end{gather}
and have the regular Abrikosov vortex which has (not only the Higgs and massive photon dressing but also) the massive magnon dressing. This is shown in Fig. \ref{3scabv} in red curves, where we have put $g'=2g$. The existence of the Abrikosov vortex in spin triplet superconductors is expected, but the Abrikosov vortex which has a massive non-Abelian magnon dressing is unexpected.

This tells that the above theory of spin triplet superconductivity actually has (not just two but) three types of vortex solutions, the Abrikosov vortex which carries the quantized magnetic flux $2\pi m/g$, the non-Abelian spin vortex which carries the quantized spin flux $2 \pi n/g'$, and the vortex made of the Higgs 
field and the massive magnon which carries no magnetic or spin flux.

Now, we discuss the monopole solution
in the spin triplet superconductors. 
Choose the monopole ansatz in the spherical coordinates $(r,\theta,\vap)$, 
\begin{gather}
\vPhi=\rho(r) \hr,
~~~\hr= \left(\begin{array}{ccc}
\sin \theta \cos \vap \\ 
\sin \theta \sin \vap \\
\cos \theta \end{array} \right), \nn\\   
A_\mu =0,   
~~~~B_\mu' =-\frac{1}{g'}(1-\cos\theta) \pd_\mu \varphi, \nn\\
\W_\mu= \frac{1}{g'} f(r)~\hr \times \pd_\mu \hr.
\label{3mans1}
\end{gather}
With this we have the following equations for the monopole, 
\begin{gather}
\ddot{\rho}+\frac{2}{r} \dot{\rho}-2 \frac{f^2}{r^2}\rho
=2\lambda\big(\rho^2-\rho_0^2 \big)\rho, \nn\\
\ddot{f}-\frac{f^2-1}{r^2}  f
-g'^2\rho^2 f =0.
\label{3meq}
\end{gather}
This is precisely the equation for 
the 'tHooft-Polyakov monopole in 
the Georgi-Glashow model \cite{tp}.

\begin{figure}
\includegraphics[height=4.5cm,width=7.5cm]
{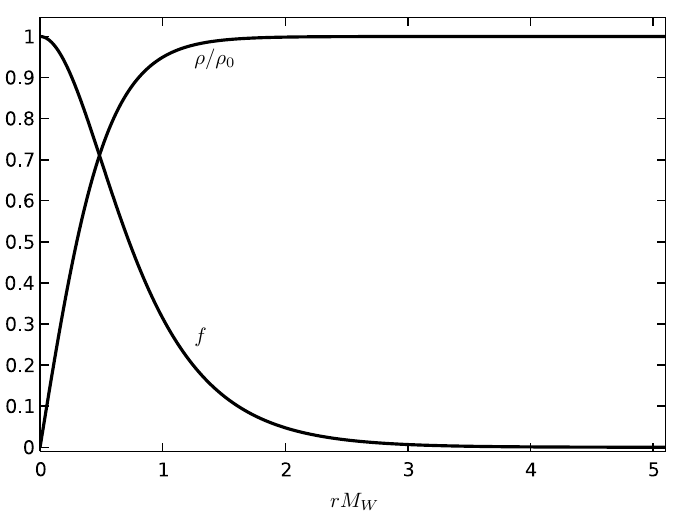}
\caption{\label{3sctpm} The 'tHooft-Polyakov type monopole solution 
which has the spin magnetic charge 
$4\pi/g'$ in the spin triplet superconductors.} 
\end{figure}

Obviously this has a singular Dirac type monopole solution
\begin{gather}
\rho=\rho_0=\sqrt{\mu^2/\lambda},~~~f=0, \nn\\
B_\mu'=-\frac{1}{g'}(1 -\cos \theta) \pd_\mu \vap,
\end{gather}
which describes the point monopole whose spin magnetic charge is given by $4\pi/g'$. Moreover, this singular monopole becomes regular with a non-trivial dressing of Higgs and the massive magnon, which implies that 
the magnon supercurrent screens the point monopole singularity at the center and regularizes it. This can be viewed as 
 evidence of the existence of 
the non-Abelian Meissner effect in spin triplet superconductors. 

Indeed we can integrate (\ref{3meq}) with 
the boundary condition 
\begin{gather}
\rho(0)=0,~~~\rho(\infty)=\rho_0, 
~~~f(0)=1,~~~f(\infty)=0.
\label{3mbcon}
\end{gather}
and find the well known 'tHooft-Polyakov monopole solution \cite{tp}. The magnonic monopole solution is shown in Fig. \ref{3sctpm}.

The 'tHooft-Polyakov monopole solution becomes the analytic Prasad-Sommerfield solution in the limit $\lambda$ goes to zero,  \cite{ps} 
\begin{gather}
\rho=\frac{1}{\tanh(r)}-\frac{1}{r},
~~~~~f=\frac{r}{\sinh(r)}.	
\end{gather}
Moreover, we can generalize the monopole to the Julia-Zee dyon which also has the spin charge \cite{jz}. Notice that these monopole and dyon are precisely the monopole and dyon that we have in the Georgi-Glashow model. But we emphasize that, unlike the 'tHooft-Polyakov monopole and Julia-Zee dyon in the Georgi-Glashow model, this monopole and dyon is not the electromagnetic monopole and dyon but the magnonic monopole and dyon which has the spin magnetic charge $4\pi/g'$.

The existence of the magnonic monopole in 
spin triplet ferromagnetic superconductors allows us to interpret the above singular magnonic vortex solution as the vortex made 
of the monopole-antimonopole pair. Actually the existence of the magnonic monopole strongly implies the existence of the quantized magnonic vortex made of 
the monopole-antimonopole pair infinitely separated apart, and the singular vortex solution could describe such a solution. 

This leads us to an important question. Does the Abrikosov vortex carrying ordinary magnetic flux also imply the existence of the electromagnetic monopole in the spin triplet superconductors? The existence of two types of conserved charges (the electric and spin charge) strongly implies the existence of two types of monopoles, the spin magnetic monopole we have discussed above and the new ordinary magnetic monopole. Indeed, a recent study shows that this is so \cite{arx1,arx2}. We will have a chance to discuss this new monopole solution in more detail later. 

\section{Comparison with Ginzburg-Landau Theory of Spin doublet Superconductivity}

It is instructive to compare the above 
theory of the spin triplet superconductivity 
with the theory of the two-gap ferromagnetic superconductivity proposed recently \cite{pla23,ap24,arx250}. Consider 
the non-Abelian SU(2)xU(1) Ginzburg-Landau theory of the ferromagnetic superconductors made of the spin doublet Cooper pair $\phi=(\phi_+,\phi_-)$ \cite{remark},
\begin{gather}
\cL =-|\cD_\mu \phi|^2 -\frac{\lambda}{2}\big(|\phi|^2
-\frac{\mu^2}{\lambda}\big)^2
-\frac14 F_\mn^2-\frac14 \G_\mn^2, \nn \\
\cD_\mu \phi 
=(D_\mu -i\frac{g}{2} A_\mu) \phi, \nn\\
D_\mu \phi=(\pd_\mu
-i\frac{g'}{2} \vsig \cdot \B_\mu \big) \phi.
\label{2lag0}
\end{gather}
where $A_\mu$ and $\B_\mu$ are 
the electromagnetic U(1) and the SU(2) gauge potentials which describe the photon and magnons, $F_\mn$ and $\G_\mn$ are 
the corresponding field strengths, $g$ and $g'$ are the coupling constants as before. Notice that here again the Cooper pair $\phi$ couples to both $A_\mu$ and $\B_\mu$, and  carries the electric and spin charge 
$g/2=2e$ and $g'$. 

With
\begin{gather}
\phi = \frac{1}{\sqrt{2}} \rho~\xi,
~~~(\xi^\dag \xi = 1), 
\label{xi}
\end{gather}
and the identity
\begin{gather}
\cD_\mu \xi= \Big[\pd_\mu -i\frac{g}{2} A_\mu 
-i\frac{g'}{2} (B_\mu' \n +\W_\mu) \cdot \vsig \Big]~\xi,  \nn\\
|\cD_\mu \xi|^2 =\frac{1}{8} (-gA_\mu+g'B_\mu')^2 
+\frac{g'^2}{4} \W_\mu^2,
\end{gather}
we can ``abelianize" the Lagrangian gauge independently \cite{pla23,ap24,arx250,remark}
\begin{gather}
\cL = -\frac12 (\pd_\mu \rho)^2
-\frac{\lam}{8}\big(\rho^2-\rho_0^2 \big)^2
-\frac14 F_\mn^2 -\frac14 {G_\mn'}^2  \nn\\
-\frac12 \big|D_\mu' W_\nu-D_\nu' W_\mu \big|^2 \nn\\	
-\frac{\rho^2}{8} \big((-gA_\mu+g'B_\mu')^2 
+2 g'^2 W_\mu^*W_\mu \big)  \nn\\
+i g' G_\mn' W_\mu^* W_\nu 
+ \frac{g'^2}{4}(W_\mu^* W_\nu 
-W_\nu^* W_\mu)^2,  \nn\\
D_\mu'=\pd_\mu +ig' B_\mu',   
~~~W_\mu =\frac{1}{\sqrt 2} (W^1_\mu +i W^2_\mu).
\label{2lag1}
\end{gather}
This tells that the Lagrangian (\ref{2lag0}) is made of two Abelian gauge potentials, the electromagnetic $A_\mu$ and the magnonic $B_\mu'$, and a complex magnon $W_\mu$ and 
the Higgs scalar field $\rho$. 

Notice, however, that the Lagrangian is not expressed by mass eigenstates. To express them in terms of mass eigenstates, 
we introduce the following photon-magnon mixing with the mixing angle $\om$ by
\begin{gather}
\left(\begin{array}{cc} \bA_\mu \\
Z_\mu  \end{array} \right)
=\frac{1}{\sqrt{g^2 +g'^2}} \left(\begin{array}{cc} 
g' & g \\ -g & g' \end{array} \right)
\left(\begin{array}{cc} A_\mu \\ B'_\mu
\end{array} \right)  \nn\\
= \left(\begin{array}{cc}
\cos \om & \sin \om \\
-\sin \om & \cos \om \end{array} \right)
\left(\begin{array}{cc} A_\mu \\ B_\mu'
\end{array} \right).
\label{mix}
\end{gather}
With this we can express the Lagrangian (\ref{2lag0}) by
\begin{gather}
\cL = -\frac12 (\pd_\mu \rho)^2
-\frac{\lam}{8}\big(\rho^2-\rho_0^2 \big)^2
-\frac14 {\bF_\mn}^2 -\frac14 Z_\mn^2 \nn\\
-\frac12 \big|(\bD_\mu +i\be\frac{g'}{g}Z_\mu)W_\nu 
-(\bD_\nu +i \be\frac{g'}{g} Z_\nu)W_\mu \big|^2  \nn\\
-\frac{\rho^2}{4} \big(g'^2 W_\mu^*W_\mu
+\frac{g^2+g'^2}{2} Z_\mu^2 \big)  \nn\\
+i \be(\bF_\mn +\frac{g'}{g} Z_\mn) 
W_\mu^* W_\nu  \nn\\
+\frac{g'^2}{4}(W_\mu^* W_\nu 
-W_\nu^* W_\mu)^2,  
\label{2lag2}
\end{gather}
where
\begin{gather}
\bF_\mn=\pd_\mu \bA_\nu-\pd_\nu \bA_\mu, 
~~~Z_\mn = \pd_\mu Z_\nu-\pd_\nu Z_\mu,  \nn\\
\bD_\mu=\pd_\mu+i \be \bA_\mu,   \nn\\
\be=\frac{gg'}{\sqrt{g^2+g'^2}}=g' \sin\om 
=g \cos\om.
\label{e}
\end{gather}	
This is the physical expression of the non-Abelian two-gap superconductivity Lagrangian (\ref{2lag0}), which tells that 
the three degrees of $\xi$ are absorbed to $Z_\mu$ and $W_\mu$ to make them massive, 
so that the theory is made of Higgs scalar $\rho$, massless and massive Abelian gauge bosons $\bA_\mu$ and $Z_\mu$, and massive complex $W_\mu$ magnon whose masses are given by 
\begin{gather}
M_H= {\sqrt \lam} \rho_0,  \nn\\ 
M_W=\frac{g'}{2} \rho_0,
~~~M_Z=\frac{\sqrt {g^2+g'^2}}{2} \rho_0.		
\end{gather} 
So it has three mass scales. 

From the physical point of view this Lagrangian looks totally different from the original Lagrangian (\ref{2lag0}). But we emphasize that the Lagrangian (\ref{2lag2}) is mathematically identical to (\ref{2lag0}), so that it retains the full non-Abelian gauge symmetry of the original Lagrangian. It is hidden, but not disappeared. 

A most important difference between this theory and the spin triplet superconductivity discussed in Section II is the photon-magnon mixing (\ref{mix}), which has deep implications. First, the two Abelian gauge bosons $\bA_\mu$ and $Z_\mu$ in (\ref{2lag2}) could naturally be interpreted to represent the real photon and the neutral magnon. This immediately tells that the massive magnon $W_\mu$ carries the electric charge as well as the spin charge, so that (just like the electron doublet) it is doubly charged. This must be clear from (\ref{2lag2}), which shows that $W_\mu$ couples to both $\bA_\mu$ and $Z_\mu$. Obviously, $W_\mu$ should carry the spin charge. What is remarkable is that it also acquires the electric charge from $\xi$ absorbing the charge carried by the unit doublet. This mixing (\ref{mix}) was absent 
in the spin triplet superconductors. So
the massive magnon in the spin triplet superconductors carries only the spin charge,
not the electromagnetic charge. 

Now, we may ask which of $\bA_\mu$ and $Z_\mu$ describes the real photon. The answer depends on whether we have a long range magnetic order in the superconductor or not. Since the long range magnetic order is an essential feature 
of the ferromagnetic superconductors, we may assume that the long range magnetic order does exist in two-gap superconductors. If so, we must identify the massless $\bA_\mu$ as 
the massless magnon which is responsible for the long range magnetic order, and identify $Z_\mu$ as the massive photon. In this case the coupling constant in front of $Z_\mu$ which couples to $W_\mu$ in (\ref{2lag2}) should become the charge of the $W_\mu$ 
magnon. This should be $2e$ because this is the charge of the spin doublet Cooper pair. 
So we must have \cite{remark} 
\begin{gather}
\be \frac{g'}{g} =2e.
\label{be}
\end{gather}  
From this we have 
\begin{gather}
g=4e,~~~g'= \sqrt{2+2\sqrt{17}}~e,  \nn\\
\tan \om= \frac{2\sqrt 2}{\sqrt{1+\sqrt{17}}},
\label{2gsc}
\end{gather} 
so that the mixing matrix is completely fixed,
\begin{gather}
\left(\begin{array}{cc}
\cos \om & \sin \om \\
-\sin \om & \cos \om \end{array} \right)  \nn\\
=\frac{1}{\sqrt{9+\sqrt{17}}}
\left(\begin{array}{cc} 
\sqrt{1+\sqrt{17}} & 2\sqrt 2 \\ 
-2\sqrt 2 & \sqrt{1+\sqrt{17}} \end{array} \right).
\label{mixf}
\end{gather}
In particular, the magnonic spin coupling $\be$ is fixed by $e$, 
\begin{gather}
\be = \frac{4\sqrt 2}{\sqrt{1+\sqrt{17}}}~e
\simeq 2.5~e,
\label{sc}
\end{gather}
so that the theory has only one coupling constant. Moreover, we have
\begin{gather}
M_W =\frac{\sqrt{1+\sqrt{17}}}{\sqrt 2}~e \rho_0, \nn\\
M_Z =\frac{\sqrt{9+\sqrt{17}}}{\sqrt 2}~e \rho_0  \simeq 1.6 M_W.
\label{2masssc}
\end{gather}
Notice that the spin coupling is stronger than the electromagnetic coupling.

With this interpretation we can express 
the Lagrangian (\ref{2lag2}) in the final form,
\begin{gather}
\cL = -\frac12 (\pd_\mu \rho)^2
-\frac{\lam}{8}\big(\rho^2-\rho_0^2 \big)^2
-\frac14 {\bF_\mn}^2  \nn\\ 
-\frac14 Z_\mn^2 
-\frac{M_Z^2}{2\rho_0^2} \rho^2 Z_\mu^2 \nn\\
-\frac12 \big|(\bD_\mu +2ie Z_\mu)W_\nu 
-(\bD_\nu +2ie Z_\nu)W_\mu \big|^2   \nn\\
-\frac{M_W^2}{\rho_0^2} \rho^2 W_\mu^*W_\mu +i (\be \bF_\mn +2e Z_\mn) W_\mu^* W_\nu   \nn\\ 
+ \frac{M_W^2}{\rho_0^2}(W_\mu^* W_\nu 
-W_\nu^* W_\mu)^2.
\label{2lag3}
\end{gather}
This tells that here the magnonic spin coupling $\be$ is completely fixed by $e$, so that the theory has only one coupling constant. This was not the case in the above spin triplet superconductors. 

Just like the spin triplet superconductivity 
the theory has two conserved currents, 
the electromagnetic and spin currents. To see this notice that we have the following equations of motion from the Lagrangian (\ref{2lag3}), 
\begin{gather}
\pd^2 \rho-\Big(2\frac{M_W^2}{\rho_0^2} W_\mu^*W_\mu 
+ \frac{M_Z^2}{\rho_0^2}~Z_\mu^2 \Big)~\rho
=\frac{\lambda}{2}\big (\rho^2 -\rho_0^2 \big)~\rho,   \nn\\	
\Big(\bD_\mu+ 2ie Z_\mu \Big) 
\Big[(\bD_\mu +2ie Z_\mu) W_\nu 
-(\bD_\nu +2ie Z_\nu) W_\mu \Big]  \nn\\
-\frac{M_W^2}{\rho_0^2} \rho^2 W_\nu 
=i W_\mu \big(\be \bF_\mn + 2e Z_\mn \big)
  \nn\\
+ 4\frac{M_W^2}{\rho_0^2} W_\mu(W_\mu^* W_\nu -W_\nu^* W_\mu), \nn\\	
\pd_\mu \Big[\bF_\mn-i\be(W_\mu^* W_\nu-W_\nu^* W_\mu) \Big] 
=4e \be~W_\mu^*W_\mu Z_\nu  \nn\\
+i\be \Big[W_\mu^* (\bD_\mu W_\nu 
-\bD_\nu W_\mu) -W_\mu (\bD_\mu W_\nu 
-\bD_\nu W_\mu)^* \Big]  \nn\\
-2e \be Z_\mu(W_\mu^*W_\nu +W_\nu^*W_\mu),  \nn\\
\pd_\mu \Big[Z_\mn -2ie(W_\mu^* W_\nu
-W_\nu^* W_\mu) \Big]   \nn\\
=\Big(\frac{M_Z^2}{\rho_0^2} \rho^2 
+8 e^2 W_\mu^*W_\mu \Big)~Z_\nu \nn\\
+2ie \Big[W_\mu^*(\bD_\mu W_\nu 
-\bD_\nu W_\mu) -W_\mu (\bD_\mu W_\nu 
-\bD_\nu W_\mu)^* \Big]   \nn\\
-4e^2 Z_\mu(W_\mu^*W_\nu +W_\nu^*W_\mu).
\label{2eom}
\end{gather}
Now, clearly the last two equations can be put in the form
\begin{gather}
\pd_\mu \bF_\mn =J_\nu^{(s)},  
~~~~\pd_\mu Z_\mn =J_\nu^{(e)},  
\end{gather}
where
\begin{gather}
J_\nu^{(s)} =4e \be W_\mu^*W_\mu Z_\nu 
+i\be \Big[\pd_\mu(W_\mu^* W_\nu-W_\nu^* W_\mu)   \nn\\
+\big(W_\mu^* (\bD_\mu W_\nu 
-\bD_\nu W_\mu) -(\bD_\mu W_\nu 
-\bD_\nu W_\mu)^* W_\mu \big) \Big]  \nn\\
-2e \be Z_\mu(W_\mu^*W_\nu + W_\nu^*W_\mu),  \nn\\ 
J_\nu^{(e)} =\Big(\frac{M_Z^2}{\rho_0^2}
\rho^2 +8 e^2 W_\mu^*W_\mu \Big)~Z_\nu  \nn\\ 
+2ie \Big[\pd_\mu(W_\mu^* W_\nu 
-W_\nu^* W_\mu)   \nn\\ 
+\big(W_\mu^*(\bD_\mu W_\nu 
-\bD_\nu W_\mu)  -W_\mu (\bD_\mu W_\nu -\bD_\nu W_\mu)^* \big) \Big] \nn\\
-4 e^2 Z_\mu(W_\mu^*W_\nu +W_\nu^*W_\mu). 
\label{2cJ}
\end{gather}
This shows that the theory has two conserved currents, the spin current $J_\mu^{(s)}$ and 
the charge current $J_\mu^{(e)}$, which correspond to the two Abelian potentials $\bA_\mu$ and $Z_\mu$.

The existence of the two conserved currents originates from the existence of two Abelian gauge symmetries of the original Lagrangian (\ref{2lag0}), the electromagnetic U(1) and the U(1) subgroup of SU(2) which leaves the Abelian direction $\n$ invariant. This must be clear in (\ref{2lag1}), which has two Abelian potentials $A_\mu$ and $B'_\mu$. So it has two conserved currents, the electromagnetic current $j_\mu^{(e)}$ of $A_\mu$ and the magnonic spin current $j_\mu^{(s)}$ of $B_\mu'$ coming from (\ref{2lag1}),
\begin{gather}
j_\nu^{(e)} 
=-2e \frac{M_Z}{\rho_0} \rho^2 Z_\nu, \nn\\  
j_\nu^{(s)} 
=\frac{M_W}{\rho_0} \Big(\frac{M_Z}{\rho_0} \rho^2 +8e W_\mu^* W_\mu \Big)~Z_\nu  \nn\\
+2i \frac{M_W}{\rho_0}  
\Big[\pd_\mu(W_\mu^* W_\nu-W_\nu^* W_\mu) \nn\\ 
+\big(W_\mu^* (\bD_\mu W_\nu-\bD_\nu W_\mu)
-W_\mu (\bD_\mu W_\nu-\bD_\nu W_\mu)^* \big) \Big] \nn\\
-2e \frac{M_W}{\rho_0} Z_\mu (W_\mu^* W_\nu +W_\nu^* W_\mu).
\label{cj}	
\end{gather} 
These two currents mix together because of the mixing (\ref{mix}), and we have the two physical currents shown in (\ref{2cJ}) after the mixing,
\begin{gather}
\left(\begin{array}{cc} J_\mu^{(s)} \\
J_\mu^{(e)}	 \end{array} \right)
= \left(\begin{array}{cc}
M_W/M_Z & \sqrt{1-(M_W/M_Z)^2} \\
-\sqrt{1-(M_W/M_Z)^2} & M_W/M_Z 
\end{array} \right)    \nn\\
\times \left(\begin{array}{cc} j_\mu^{(e)} \\ j_\mu^{(s)} \end{array} \right).
\label{Jmix}
\end{gather} 
The existence of the charge current in 
the non-Abelian superconductors is natural, but the existence of the conserved spin current in the spin doublet superconductors has not been well appreciated so far. 

This tells two things. First, both $J_\mu^{(s)}$ and $J_\mu^{(e)}$ contain the electromagnetic current $j_\mu^{(e)}$ of $A_\mu$ and the spin current $j_\mu^{(s)}$ of $B_\mu'$. Second, both (in particular, $J_\mu^{(s)}$) contain the spin current of the Cooper pair and the spin  current of the magnon. From this we could expect the conversion of the spin current and charge current as well as the interconversion between the spin current of the Cooper pair and the spin current of the magnon. For example, the conserved spin current $J_\mu^{(s)}$ is made of the electromagnetic current $j_\mu^{(e)}$ and the magnonic spin current $j_\mu^{(s)}$. As far as we understand, there has been no theory of non-Abelian superconductivity which has these properties.  

There are a few points to be mentioned before 
we leave this section. First, here again the Higgs mechanism takes place without any spontaneous symmetry breaking. The mass generation of the photon $Z_\mu$ and charged magnon $W_\mu$ in (\ref{2lag2}) comes from the non-vanishing vacuum value of the density of the electron pair $\rho$, which does not break any symmetry, spontaneous or not. In fact, here $\langle \xi \rangle$ could still fluctuate with the Higgs mechanism even at the vacuum \cite{pla23,ap24,arx250}.

Second, just like the theory of the spin triplet superconductivity, this is a theory 
in which the spin-spin interaction is described by the exchange of the messenger particle, the magnons. 

Third, the Lagrangian (\ref{2lag0}) is precisely the Weinberg-Salam Lagrangian of the standard model in high energy physics, which was originally proposed by Weinberg to unify the electromagnetic and weak interactions \cite{wein}. So it is really remarkable that exactly the same Lagrangian could describe the non-Abelian superconductivity in two-gap ferromagnetic superconductors. 

Of course, the Lagrangian (\ref{2lag0}) has 
a totally different meaning here. For instance, in the standard model the massless $\bA_\mu$ describes the real massless photon, but here it describes the massless magnon. Moreover, the massive photon $Z_\mu$ describes the neutral weak boson. As importantly, the energy scale in two theories is totally different. In fact, the Higgs vacuum value in the standard model is of the order of 100 GeV, but in condensed matter physics the Higgs vacuum is supposed to be 
of the order of meV, different by 
the factor $10^{14}$. So they describe totally different physics in totally different surroundings. 

\section{Topological Objects in Spin Doublet Superconductors}  

Just like the spin triplet superconductors, the spin doublet ferromagnetic superconductors have the same types of topological objects, two quantized vortices and one monopole \cite{pla23,ap24}. This is because the two theories are based on the same non-Abelian SU(2)xU(1) gauge symmetry. But the physical contents of these topological objects are different from those in the spin triplet superconductors.

To show this, we choose the string ansatz 
in the cylindrical coordinates $(r,\vap,z)$ \cite{pla23,ap24,epjc24}, 
\begin{gather}
\phi= \frac{1}{\sqrt 2} \rho(r)~\xi,
~~~\xi= \left(\begin{array}{cc}
-\sin \dfrac{\al(r)}{2}~\exp (-in \vap) \\
\cos \dfrac{\al(r)}{2} \end{array} \right), \nn\\
~~~A_\mu = \frac{m}{g} A(r)~\pd_\mu \vap,  \nn\\
\hB_\mu=n \frac{1}{g'} \big(B(r)+1
-\cos\alpha(r)\big) \pd_\mu \vap~\n  -\frac{1}{g'}~\n \times \pd_\mu \n,  \nn\\
\W_\mu= \frac{1}{g'} f(r)~\n \times \pd_\mu \n,   \nn\\
\n =-\xi^\dagger \vsig \xi
=\left(\begin{array}{ccc}
\sin \al(r) \cos n\vap \\ 
\sin \al(r) \sin n\vap \\
\cos \al(r)  \end{array} \right),
\label{2sans0}
\end{gather} 
where $m$ and $n$ are integers which represent the winding numbers of the $\pi_1(S^1)$ topology of U(1) and U(1) subgroup of SU(2). In terms of the physical field, the ansatz becomes \cite{epjc24}
\begin{gather}
\rho=\rho(r), 
~~~~\bA_\mu = \be \Big(\frac{m}{g^2} A
+\frac{n}{g'^2} B \Big)~\pd_\mu \vap,    \nn\\
Z_\mu=-\frac{1}{\sqrt{g^2+g'^2}} 
\big(m A -n B \big)~\pd_\mu \vap,  \nn\\
\W_\mu= \frac1{g'} f ~\n \times \pd_\mu \n.
\label{2sans1}
\end{gather}
So, when $mA=nB$ the ansatz describes the magnon vortex made of 
$\bA_\mu$,
\begin{gather}
\bA_\mu= \frac{m}{\be} A~\pd_\mu \vap,
~~~Z_\mu=0.
\label{2magv}
\end{gather}
But when $mA/g^2+nB/g'^2=0$, the ansatz describes the magnetic vortex made of 
the massive photon $Z_\mu$ 
\begin{gather}
\bA_\mu =0,   \nn\\
Z_\mu =n \frac{\sqrt{g^2+g'^2}}{g'^2} B~\pd_\mu \vap =\frac{n}{e} Z~\pd_\mu \vap,
\label{2abv}
\end{gather}
where $Z=B/2$. This suggests that the spin doublet ferromagnetic superconductors have two types of vortices. 

\begin{figure}
\includegraphics[height=4.5cm, width=8cm]{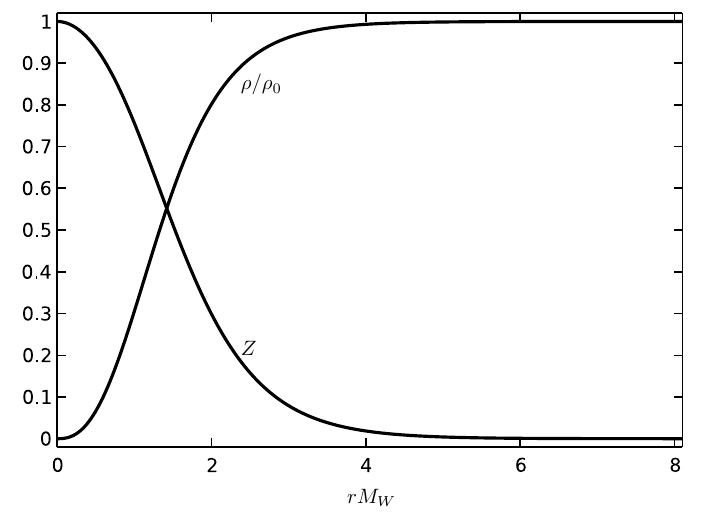}
\caption{\label{2scabv} The Abrikosov vortex solution in the spin doublet superconductors made of the Higgs scalar and massive photon which has the quantized magnetic flux $2\pi/e$.}
\end{figure}

To show this notice that, when $\al$ becomes 
a constant, we have the following string equations of motion from the ansatz (\ref{2sans1}),
\begin{gather}
\ddot \rho+\frac{\dot \rho }{r}
-\frac{n^2 f^2 \sin^2 \alpha 
+(mA-nB)^2}{4 r^2}~\rho
=\frac{\lambda}{2}\rho(\rho^2-\rho_0^2),  \nn\\
n f\rho (mA +n)\sin\alpha =0, \nn\\	
m \big(\ddot A-\frac{\dot A}{r} \big) 
=-\frac{g^2}{4} \big(mA-nB \big) \rho^2, \nn\\	
n \big(\ddot B-\frac{\dot B}{r} \big)
=-\frac{g'^2}{4} \big(mA-nB \big) \rho^2, \nn\\
\sin \al \Big(\ddot f -\frac{\dot{f}}{r} 
-\frac{g'^2}{4} \rho^2f \Big) =0,  \nn\\	
n^2 \Big((B+1)\dot f-f \dot B \Big) \sin \al
=0.
\label{seq}	
\end{gather}
Now, consider the electromagnetic vortex made of $Z_\mu$ first. With $mA/g^2+nB/g'^2=0$ and $\al=0$, the above equation reduces to \cite{epjc24}
\begin{gather}
\ddot \rho+\frac{\dot \rho}{r}
-n^2 \frac{(g^2+g'^2)^2}{g'^4}
\frac{Z^2}{r^2} \rho 
=\frac{\lambda}{2}\rho(\rho^2-\rho_0^2),  \nn\\
\ddot Z-\frac{\dot Z}{r} 
-\frac{g^2+g'^2}{4} \rho^2 Z =0.
\label{zseq}
\end{gather}
Obviously this is the equation of the well known Abrikosov vortex we have in the ordinary superconductors \cite{abri,no}. We can solve (\ref{zseq}) with the boundary condition 
\begin{gather}
\rho(0)=0,~~~\rho(\infty)=\rho_0,   
~~~Z(0)=1,~~~Z(\infty)=0,
\end{gather} 
and obtain the Abrikosov vortex solution which has the quantized magnetic flux $2\pi/e$,
\begin{gather}	
\Phi=\oint_{r=0}^{r=\infty} Z_\mu dx^\mu
= \frac{2\pi n}{e},~~~Z_\mu=-\frac{n}{e} Z~\pd_\mu \vap.
\label{2abv}
\end{gather}
The solution for $n=1$ is shown in Fig. \ref{2scabv}.

\begin{figure}
\includegraphics[height=4.5cm, width=8cm]{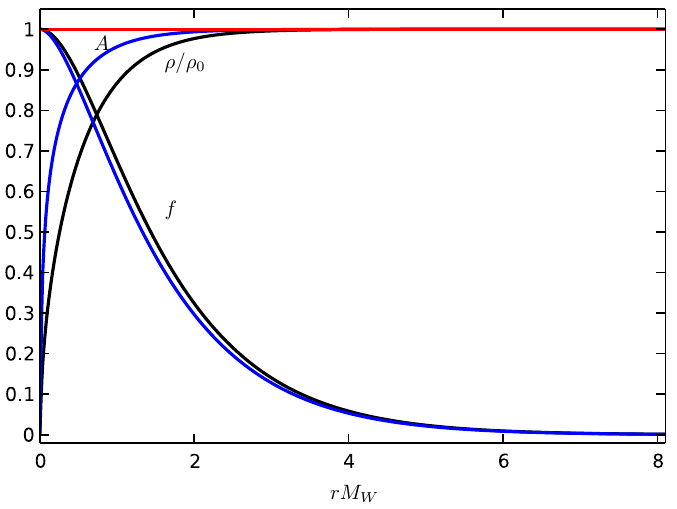}
\caption{\label{2scnav} The quantized spin vortex carrying the spin flux $2\pi/\be$ in the spin doublet superconductors shown in the red line $A=1$, which can have the Higgs and massive magnon dressing shown in black curves.
The blue curves represent the solution 
with $n=0$. Notice that the Higgs and massive magnon configuration can also become an independent vortex solution without the singular spin vortex.}
\end{figure}

Next, consider the magnonic vortex made of 
$\bA_\mu$. With $B=-1$ and $\al=\pi/2$ 
we have $mA+n=0$, so that the equation 
(\ref{seq}) becomes 
\begin{gather}
\ddot \rho+\frac{\dot \rho}{r}
-\frac{n^2}{4} \frac{f^2}{r^2}~\rho
=\frac{\lambda}{2}(\rho^2-\rho_0^2) \rho,  \nn\\
\ddot f -\frac{\dot f}{r} 
-\frac{g'^2}4 \rho^2 f =0.
\label{sveq}
\end{gather}	
Clearly this has the naked singular string solution given by
\begin{gather}
\rho=\rho_0,~~~f=0,~~~A =-\frac{n}{m} =1, 
\label{ssv}
\end{gather}
which has the quantized spin flux generated by $\bA_\mu$,
\begin{gather}	
\Phi=\oint \bA_\mu dx^\mu 
=\frac{2\pi m}{\be},
~~~\bA_\mu= \frac{m}{\be} \pd_\mu \vap.
\label{sf}
\end{gather}
Moreover, we can solve (\ref{sveq}) with the boundary condition
\begin{gather}
\rho(0)=0,~~~\rho(\infty)=\rho_0,  
~~~f(0)=1,~~~f(\infty)=0,
\end{gather}
and find the dressed spin vortex dressed by the Higgs and massive magnon,
which is shown in Fig. \ref{2scnav} in black curves.

Furthermore, we can also have the vortex 
made of the Higgs and massive magnon without $\bA_\mu$. To see this, notice first that we can gauge away $\bA_\mu$ with 
a singular gauge transformation, and show 
that the equation (\ref{sveq}) remains 
the same after the gauge transformation. This means that the vortex made of the Higgs and the massive magnon becomes a solution of 
the spin doublet superconductors without $\bA_\mu$. Second, when $n=0$ (and $\bA_\mu=0$), the Higgs and the massive magnon equations in (\ref{sveq}) decouple but have the solutions shown in the blue curves in Fig. \ref{2scnav}. 
	
This tells that the spin doublet non-Abelian superconductors also have three types of quantized vortices, the one which has the quantized spin flux, the one which has the quantized magnetic flux, and the one which 
has no flux at all.

Now, we discuss the monopole solution. 
Choose the monopole ansatz in the spherical coordinates $(r,\theta,\vap)$ given 
by \cite{plb97}  
\begin{gather}
\rho =\rho(r),~~~\xi
=i\left(\begin{array}{cc} \sin (\theta/2)~e^{-i\varphi} \\
- \cos(\theta/2) \end{array} \right),  \nn\\
A_\mu =-\frac{1}{g}(1-\cos\theta) \pd_\mu \varphi, \nn\\
\B_\mu= \frac{1}{g'}(f(r)-1)~\hr \times \pd_\mu \hr,
\label{2mans1}
\end{gather}
which can be expressed in terms of the physical fields by
\begin{gather}
\rho =\rho(r),
~~~\bA_\mu =-\frac{1}{\be}(1-\cos\theta) \pd_\mu \varphi, \nn\\
Z_\mu =0,~~~\W_\mu= \frac{1}{\be} f(r)~\hr \times \pd_\mu \hr.
\label{2mans2}
\end{gather}
With this we have the equations of motion 
\begin{gather}
\ddot{\rho}+\frac{2}{r} \dot{\rho}-\frac{f^2}{2r^2}\rho
=\frac {\lambda}{2}\big(\rho^2
-\rho_0^2 \big)\rho, \nn\\
\ddot{f}-\frac{f^2-1}{r^2}f
-\frac{g'^2}{4}\rho^2 f =0.
\label{deq1} 
\end{gather}
Obviously this has a singular solution
\begin{gather}
\rho=\rho_0=\sqrt{2\mu^2/\lambda},~~~f=0, \nn\\
\bA_\mu=-\frac{1}{\be}(1 -\cos \theta) \pd_\mu \vap,
\end{gather}
which describes the Dirac type point 
monopole carrying the spin magnetic charge 
$4\pi/\be$. 

\begin{figure}
\includegraphics[height=4.5cm, width=7.5cm]{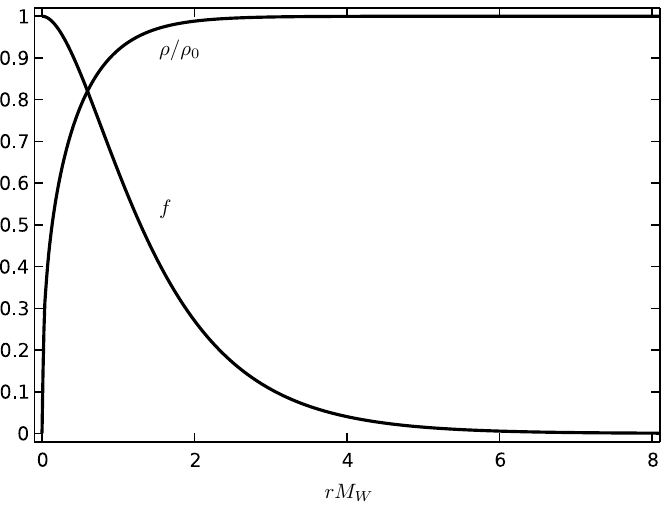}
\caption{\label{2sccmm} The Cho-Maison monopole which has the W magnon and Higgs scalar dressings carrying the spin magnetic charge $4\pi/\be$ in spin doublet ferromagnetic superconductors.}
\end{figure}

Remarkably, this monopole naturally admits 
a non-trivial dressing of Higgs and $W$ magnon. Indeed we can integrate (\ref{deq1}) with the boundary condition 
\begin{gather}
\rho(0)=0,~~~\rho(\infty)=\rho_0, 
~~~f(0)=1,~~~f(\infty)=0.
\label{mbcon}
\end{gather}
and find the dressed monopole solution which can be viewed as a hybrid between the Dirac and 'tHooft-Polyakov monopoles. This is the Cho-Maison monopole shown in Fig. \ref{2sccmm}, first discovered in the standard model \cite{plb97}. 

Just as in the spin triplet superconductors, the existence of the Cho-Maison monopole in 
the spin doublet ferromagnetic superconductors allows us to interpret the singular magnonic vortex solution shown in Fig. \ref{2scnav} as the Cho-Maison monopole-antimonopole pair infinitely separated apart. The only point is that the spin magnetic flux of the Cho-Maison monopole-antimonopole pair should be $4\pi/\be$, but the spin vortex is 
$2\pi n/\be$ as shown in (\ref{sf}). So, 
with this in mind, we can indeed interpret 
the magnonic spin vortex solution as the vortex made of the Cho-Maison monopole-antimonopole pair.  

It should be emphasized that the Cho-Maison monopole in the standard model is a real electromagnetic monopole which has the real magnetic charge $4\pi/e$, but in the spin doublet superconductors the Cho-Maison monopole is not electromagnetic but spin magnetic monopole which has the spin magnetic charge $4\pi/\be$.

This raises the same question that we had in the spin triplet superconductors. Since we have two types of the vortex solutions, can 
we also have the electromagnetic monopole which has the magnetic charge $4\pi/e$ in 
the spin doublet superconductors? The recent study tells that indeed this is so, and the
spin doublet superconductors may have two 
types of monopoles, the Cho-Maison type 
spin magnetic monopole discussed above and 
a totally new type of electromagnetic 
monopole which carries the magnetic charge $4\pi/e$ \cite{arx1,arx2}. This is a really interesting development. 

\section{Comparison with Non-Abelian Magnon Spintronics}

It has been well known that the magnon spintronics is closely related to the non-Abelian superconductivity \cite{chumak,
jmmm,hb}, which has made the superconducting spintronics an important part of spintronics \cite{lind,esch}. In both cases the non-Abelian spin structure and the magnon spin-spin interaction play the central role, and the long range magnetic order becomes an essential feature of the theories. But these results were mostly the experimental observations, and the theoretical explanation 
why this is so has been lacking. This 
situation has changed now, and a recent study shows that the theory of non-Abelian superconductivity and that of the spintronics could have a deep connection because mathematically they are described by the same theory \cite{arx251}.  

To discuss this point in more detail, we 
start from the non-Abelian spin doublet ferromagnetic superconductivity described by the Lagrangian (\ref{2lag0}), where the Higgs doublet $\phi$ is interpreted as the spin doublet Cooper pair. Clearly, if we interpret the Higgs doublet as the charged spinon which represents the electron (not as the spin doublet Cooper pair), the Lagrangian (\ref{2lag0}) could describe the physics of the electron spintronics \cite{arx250,arx251}. 

Indeed, with this interpretation of the Higgs doublet in (\ref{2lag0}), the Lagrangian can describe the non-Abelian gauge theory of electron spintronics, with the SU(2) gauge field $\B_\mu$ as the non-Abelian magnon. 
In this case the Abelian $A_\mu$ describes 
the electromagnetic interaction and 
the non-Abelian $\B_\mu$ describes the spin-spin interaction of the electron. 

This changes the details of the theory. 
The main change is that in the spin doublet superconductors the electric charge of 
the Cooper pair is $2e$, but in the electron spintronics the electron charge is $e$. So we must have $g/2=2e$ in the superconductor, but $g/2=e$ in the electron spintronics. This changes the gauge boson masses in the electron spintronics. In the electron spintronics we have  
\begin{gather}
g=2e,~~~g'=\frac{\sqrt{1+\sqrt{17}}}{\sqrt 2} e,   \nn\\
\tan \om= \frac{2\sqrt 2}{\sqrt{1+\sqrt{17}}}
\label{2gest}
\end{gather} 
so that 
\begin{gather}
M_W =\frac{\sqrt{1+\sqrt{17}}}{2\sqrt 2} e \rho_0, \nn\\
M_Z =\frac{\sqrt{9+\sqrt{17}}}{2\sqrt 2} e \rho_0  \simeq 1.6 M_W.
\label{2massest}
\end{gather}
This should be compared with (\ref{2gsc})
and (\ref{2masssc}). This tells that, although the coupling constants $g$ and $g'$ and the masses $M_W$ and $M_Z$ change, the mixing angle and the ratio between two masses remain the same in the electron spintronics.  

With this change, the Lagrangian can describe all known important properties of the electron spintronics, in particular the existence of the long range magnetic order guaranteed by the massless magnon $\bA_\mu$, the existence of conserved charge and spin currents, and the mixing of the two conserved currents which allows the interconversion between the two currents. This tells that the same Lagrangian (\ref{2lag0}) can describe not only the physics of the spin doublet ferromagnetic superconductivity but perhaps more importantly the physics of the electron spintronics. This assures that the two theories are not just closely related but almost identical to each other \cite{arx250,arx251}. This is really remarkable. 

This has an important implication on 
the electron spintronics. If the above observation is correct, the photon-magnon mixing becomes inevitable in the electron spintronics. And indeed, this might be precisely what we need, because it is well 
known that in the electron spintronics we 
have the conversion between charge and spin currents \cite{chumak,jmmm,hb}. And this conversion can be due to the photon-magnon mixing as we have explained in (\ref{Jmix}). 

The above discussion suggests that exactly the same connection can be established 
between the spin triplet superconductivity 
and the spin triplet magnon spintronics. 
There have been indications in the literature 
that the two theories are closely 
connected \cite{lind,esch}. And our Lagrangian (\ref{3lag0}) can also describe the spin triplet spintronics. Here the connection is more transparent, because in both cases the Higgs triplet $\vPhi$ represents the spin triplet Cooper pair.   

So, here the two theories really become identical. Moreover, the two phenomena 
can often take place in the same materials. The spin triplet ferromagnetic materials have the spin triplet spintronic phenomena, and 
the spin triplet spintronic materials have 
the spin triplet superconducting phenomena \cite{lind,esch}. This would be impossible 
in the spin doublet superconductors, because here the Higgs doublet has the charge $2e$, while the Higgs doublet in the electron spintronics carries the charge $e$.

Moreover, here things become simpler. Unlike the spin doublet superconductors, there is 
no photon-magnon mixing in the spin triplet superconductors. But we still have two conserved currents, charge and spin currents. And, of course, the conserved spin current plays an important role in both 
the non-Abelian superconductors and 
the spintronic materials.  

From this we may conclude that the underlying physics of the spin doublet and triplet superconductivity is really the same as the underlying physics of the electron spintronics and the spin triplet magnon spintronics. Both are described by the non-Abelian magnon gauge interaction. The only difference is the interpretation.

\section{Discussions}

In this paper we have shown how to generalize the Abelian Ginzburg-Landau theory of superconductivity to a non-Abelian spin triplet Ginzburg-Landau theory of superconductivity which has the SU(2)xU(1) gauge interaction. This is a straightforward generalization of the non-Abelian Ginzburg-Landau theory of spin doublet ferromagnetic superconductivity proposed 
recently \cite{pla23,ap24,arx250}.

The theory has the following important features. First, just as in the spin doublet ferromagnetic superconductivity, it has 
the long range magnetic order guaranteed by the existence of the massless magnon.
Second, it has two conserved currents, 
the charge and spin currents.
Third, it has the non-Abelian Meissner effect which screens the spin flux, as well as the ordinary Meissner effect which screens the magnetic flux. But unlike the spin doublet ferromagnetic superconductivity, this happens without the photon-magnon mixing. So the spin triplet superconductors have no interconversion of the two conserved currents.

Moreover, just like the spin doublet ferromagnetic superconductors, the spin triplet superconductors have interesting topological objects. They have three types of vortices, the one carrying the quantized  spin flux $2\pi n/g'$, the one carrying the quantized magnetic flux $2\pi m/g$, and the one carrying no flux at all. Moreover, they have two types of monopoles, the magnonic monopole of the Cho-Maison type which has the spin magnetic charge $4\pi/g'$ and the electromagnetic monopole which has the magnetic charge $4\pi/g$. The existence of these topological objects comes from the two $\pi_1(S^1)$ vortex topology coming from two U(1) gauge symmetries, the electromagnetic U(1) and 
the Abelian U(1) subgroup of SU(2), and 
the two monopole topology of the U(1) and SU(2) gauge symmetry. 

An important aspect of the spin triplet ferromagnetic superconductivity is that 
(just like the spin doublet superconductivity) it is described by a non-Abelian magnon gauge interaction which describes the spin-spin interaction in terms of the exchange of the messenger particle. This is remarkable for two reasons. First, the non-Abelian gauge interaction has rarely been used in condensed matter physics till recently, although the non-Abelian condensed matters have become very popular. Indeed, only recently this type of non-Abelian gauge interaction has been proposed to describe the non-Abelian condensed 
matters \cite{pla23,ap24,arx250,zarkim}. Second, so far the spin-spin interaction 
has always been viewed as an instantaneous action at a distance. This has been strange, because this is against the causality. And this theory treats the spin-spin interaction as a normal interaction generated by the exchange of messenger particles.  

In this paper we have assumed the existence of the magnonic gauge interaction in the spin triplet superconductors. As we have mentioned in the introduction, however, 
some of the spin triplet superconductors 
(in particular non-ferromagnetic superconductors) may have no magnon gauge interaction. In this case the SU(2) symmetry becomes global, and the theory becomes a theory of non-Abelian spin triplet superconductivity which has a global SU(2) symmetry. This is an interesting theory worth pursuing further \cite{prb05,prb06,epjb08}.  

Do we really have the spin triplet superconductors which can be described by 
the above theory? Only experiment can tell 
the answer. We have proposed the theory, and the experiment should check the validity of the theory. It would be interesting to see 
if we can establish the non-Abelian Meissner effect and non-Abelian magnon spin current in the spin triplet superconductors experimentally.   

In this paper we have discussed the effective theory of the spin triplet 
superconductors, but certainly one could ask what is the microscopic origin of 
this theory. In particular, one could ask what is the pairing mechanism of the spin triplet Cooper pair. This is a legitimate question which needs to be discussed
in detail. Our purpose of this paper 
is to discuss the physics of the effective theory of the spin triplet superconductivity without touching the complicated issue on the pairing mechanism. 

{\bf ACKNOWLEDGEMENT}

~~~LZ and PZ are supported by the National Key R\&D Program of China (No. 2024YFE0109802) and National Natural Science Foundation of China (Grant No. 12175320 and No. 12375084). YMC 
is supported in part by the President's Fellowship Initiative of Chinese Academy of Sciences (Grant No. 2025PD0115), the National Research Foundation of Korea funded by 
the Ministry of Education (Grant 2022-R1A2C1006999), and by Center for Quantum Spacetime, Sogang University, Korea.


\begin{references}
\bibitem{remark} Notice that in Refs. [9] 
and [10] the magnon is called the ``s-boson", because the authors were advised not to call this the magnon. These references also contain a few typographical mistakes which are corrected in this paper. For example, $g=2e$ should change to $g=4e$.
\end{references}
\end{document}